\documentclass[10pt,oneside,letter]{article}
\usepackage[font=footnotesize,labelfont=bf]{caption}
\usepackage{tabularx}
\usepackage{color}
\usepackage{amsmath,amsfonts,amssymb,amscd}
\usepackage{graphicx}
\usepackage{cite}
\usepackage{multirow}
\usepackage{booktabs}
\usepackage{paralist}
\usepackage{ulem}
\usepackage{enumitem}
\usepackage{multicol}
\usepackage{wrapfig,lipsum,booktabs}
\usepackage{epstopdf}
\usepackage{subfigure}
\usepackage{gensymb}
\usepackage{tikz}
\usepackage{bm}
\usepackage{mathtools}

\usepackage{steinmetz}
\usepackage{soul}
\usepackage{wrapfig}
\usepackage[ampersand]{easylist}
\usepackage{subfig}
\usepackage[english]{babel}
\usepackage{blindtext}
\usepackage{wrapfig,lipsum,booktabs}
\usepackage[compact]{titlesec}
\titlespacing{\section}{0pt}{*0}{*0}
\titlespacing{\subsection}{0pt}{*0}{*0}
\titlespacing{\subsubsection}{0pt}{*0}{*0}
\usepackage[left=2cm,top=2cm,right=2cm,bottom=1.5cm]{geometry}
\setdescription{leftmargin=\parindent,labelindent=\parindent}
\setdescription{leftmargin=0.5\parindent}
\usepackage{setspace}
\usepackage{scalerel}

\graphicspath{{./figures/}}
\usepackage{fancyhdr}
\usepackage{lipsum}
\let\OLDthebibliography\thebibliography
\renewcommand\thebibliography[1]{
	\OLDthebibliography{#1}
	\setlength{\parskip}{0pt}
	\setlength{\itemsep}{0pt plus 0.3ex}
}

\begin{document}	
\title {\large  \bf \vspace{-12ex}
	Performance Limits of Differential Power Processing \vspace{-4ex} 
}
\date{}
\author{\small \vspace{-2.8ex} Ping Wang$^1$, Robert C. N. Pilawa-Podgurski$^2$, Philip T. Krein$^3$, Minjie Chen$^1$\\ \small $^1$Princeton University, $^2$University of California, Berkeley, $^3$University of Illinois at Urbana-Champaign}
\maketitle
\thispagestyle{empty}
\pagestyle{empty}

\renewcommand{\figurename}{Fig.}
\vspace{-4ex}
\noindent {\bf Abstract}: This paper investigates the performance limits of differential power processing (DPP) and presents quantitative and systematic design guidelines for the selection and comparison of DPP topologies. A stochastic model is developed to evaluate the expected power losses of a variety of DPP topologies with probabilistic load distribution. The expected losses of several DPP topologies are derived and compared against traditional dc-dc converters to reveal their performance limits. The impacts of the load distribution and load scale on the expected losses are investigated. The theoretical models are verified with SPICE simulations and experimental results.

\section{\large Introduction}\label{sec:Intro}
\begin{wrapfigure}{r}{0.45\textwidth}
	\vspace{-35pt}
	\begin{center}
		\includegraphics[width=0.45\textwidth]{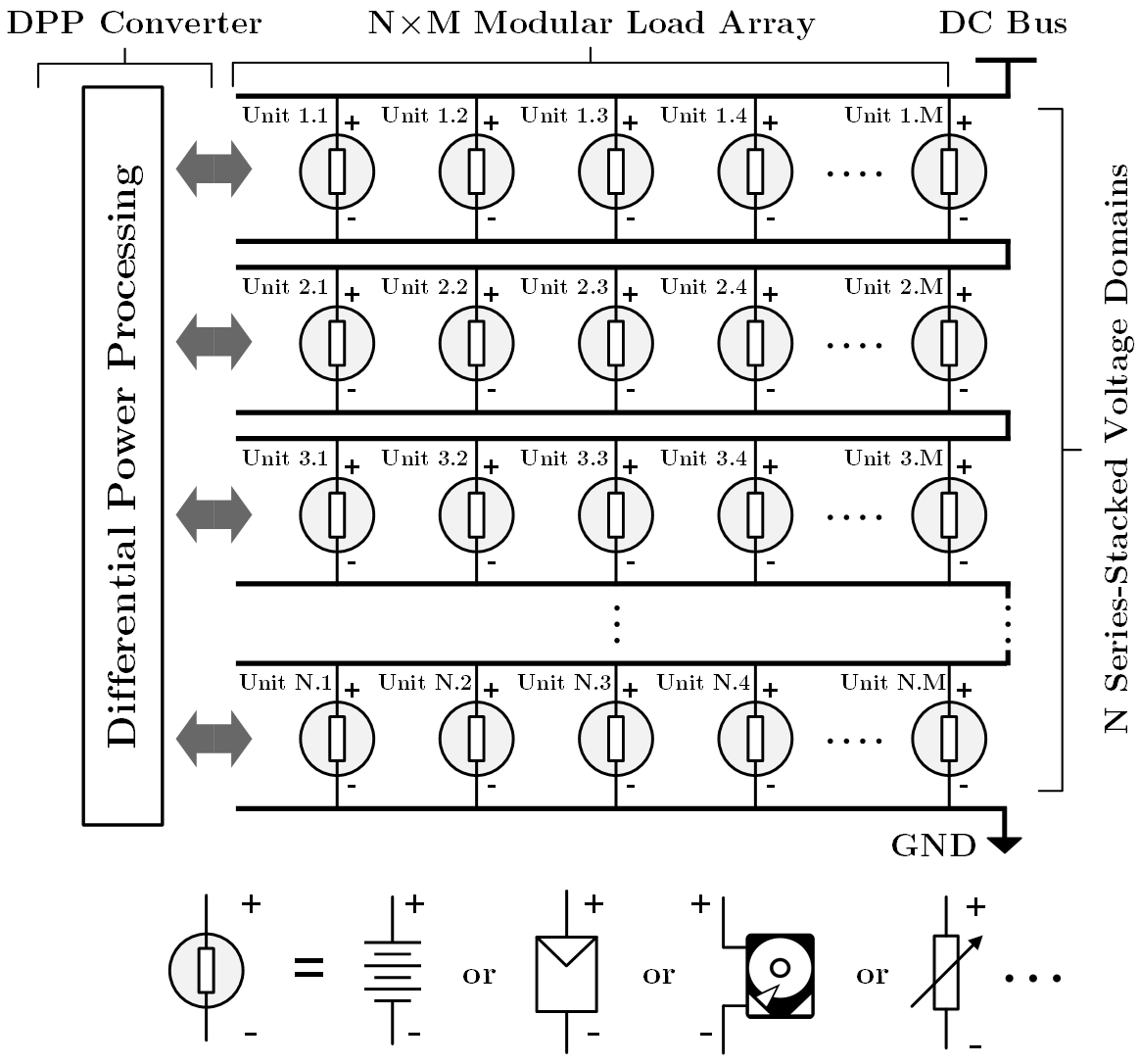}
		\caption{A $N \times M$ differential power processing system with $N$ series-stacked voltage domains, each comprising $M$ modular loads. The modular load units can be battery cells, PV panels, or hard disk drives (HDD).}
		\label{fig:DPP4Array}
	\end{center}
	\vspace{-30pt}
\end{wrapfigure}

Differential power processing (DPP) has been proved effective in many applications including solar photovoltaics, battery management systems, and computers on server racks (Fig.~\ref{fig:DPP4Array})~\cite{Pradeep13, Pilawa16, Kutkut94, Schmidt93, Brainard95}. DPP converters only process the differential power and can greatly reduce the power conversion stress and losses. Various DPP topologies have been explored, with tradeoffs in efficiency, size, cost, and control complexity. Empirical work has been done to compare DPP topologies with traditional dc-dc topologies using numerical and SPICE simulations \cite{Pilawa16}. An analytical model providing generalized design guidelines for DPP topologies under a set of rigorous assumptions is needed and is the main focus of this paper.

This paper systematically investigates the performance limits of differential power processing. A performance scaling factor, $\mathcal{S}(\bullet)$, is introduced to describe how the performance of a DPP converter changes as the system size scales up. The model has a minimum set of assumptions, offers rich design insights, and is verified with SPICE simulations and experimental results.  Extended theoretical derivations, modeling details, and experimental results will be provided in the full paper.

\section{\large Stochastic Loss Model and Scaling Factor of DPP Topologies }\label{sec:Model}
\begin{figure}[t]
	\begin{center}
		\hspace{-17pt}
		\subfigure[]{\includegraphics[height=0.18\textheight]{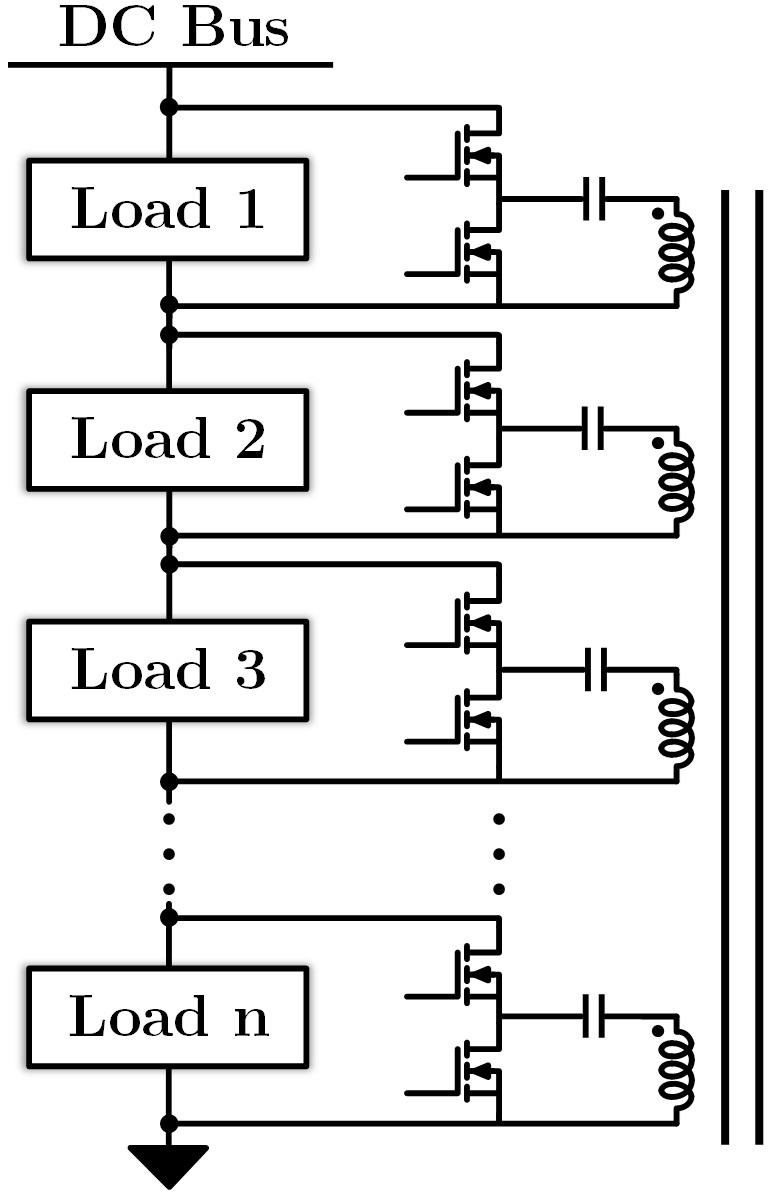}}
		\hspace{6pt}
		\subfigure[]{\includegraphics[height=0.18\textheight]{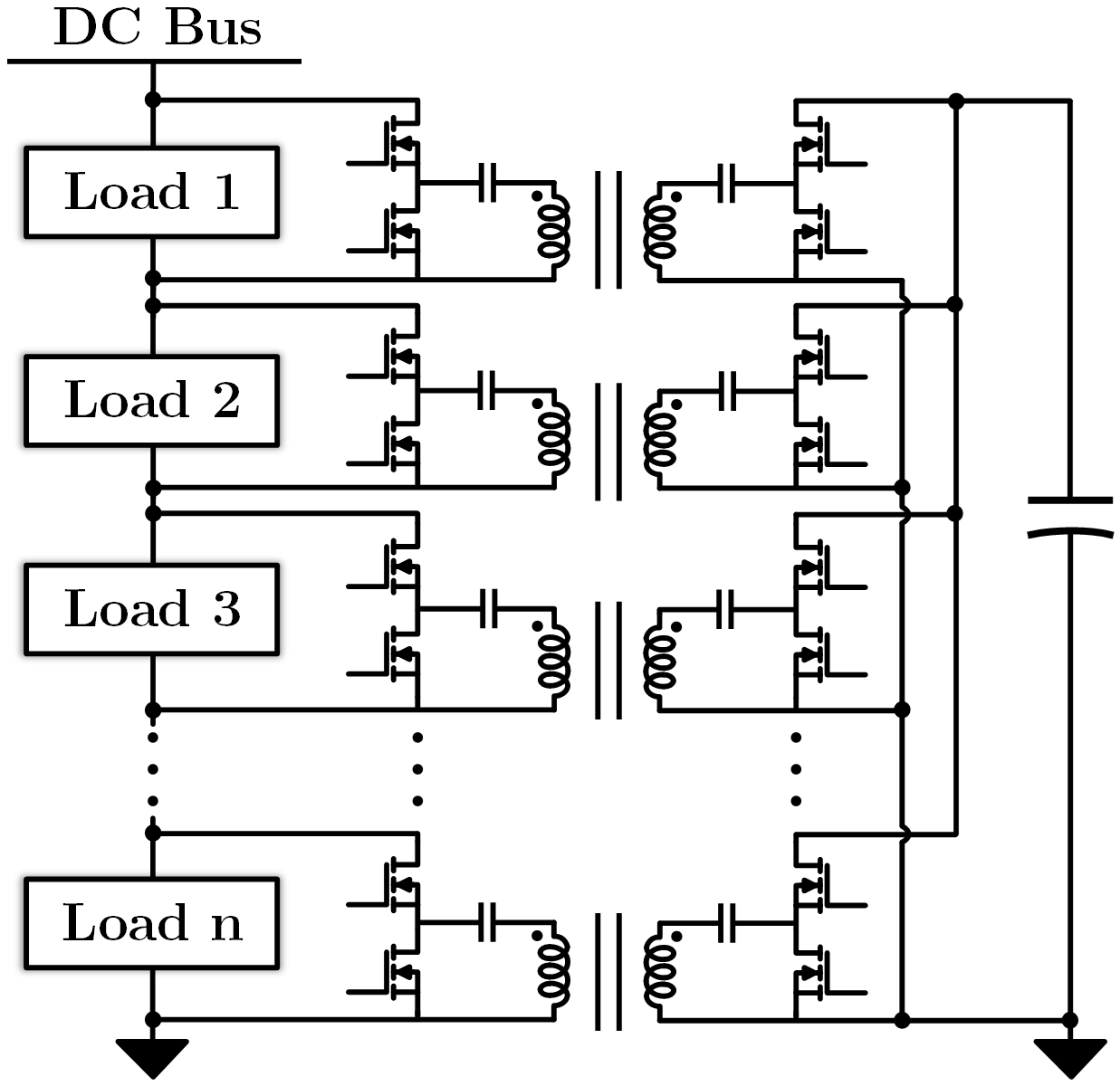}}
		\hspace{6pt}
		\subfigure[]{\includegraphics[height=0.18\textheight]{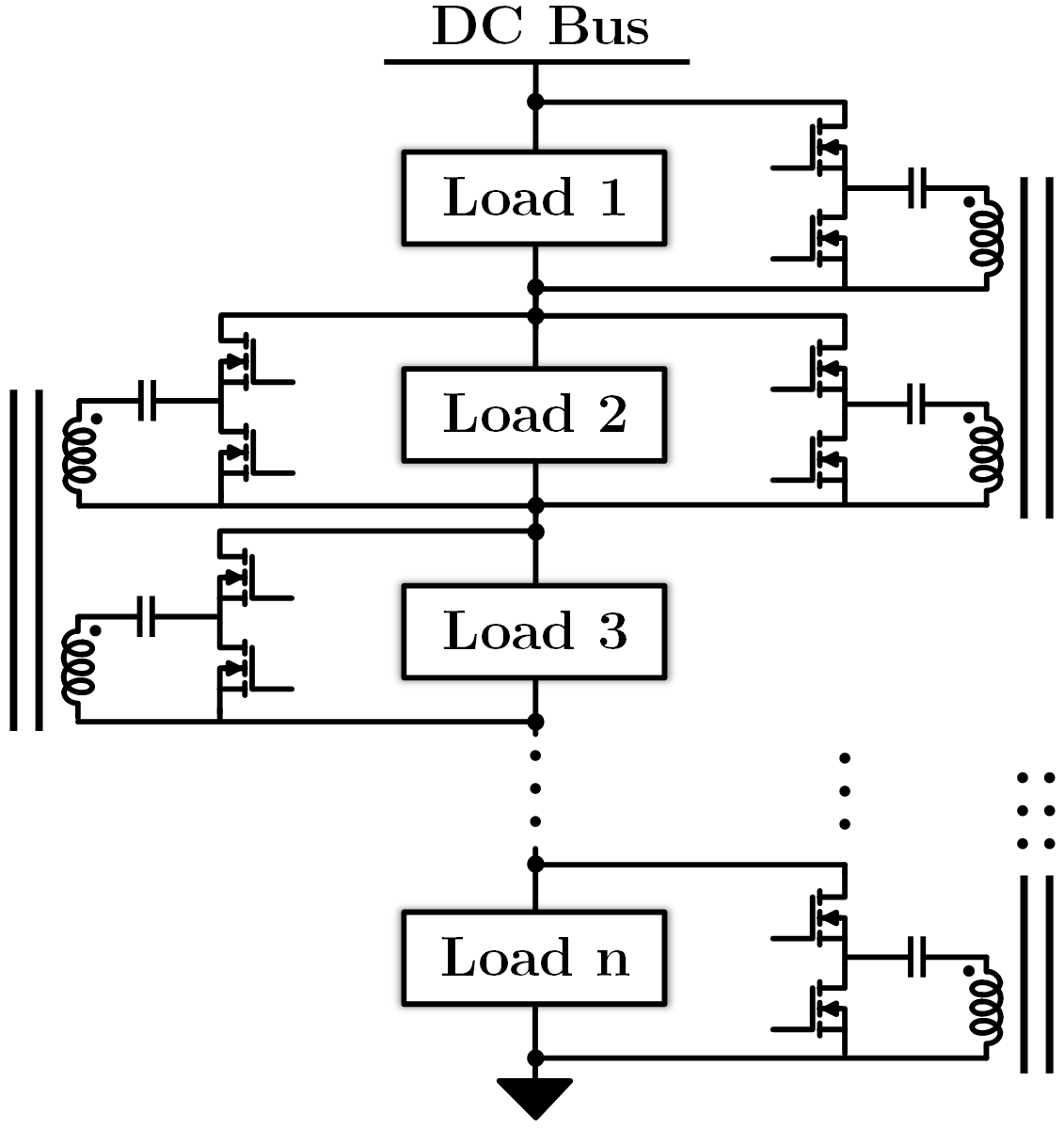}}
		\hspace{16pt}
		\subfigure[]{\includegraphics[height=0.18\textheight]{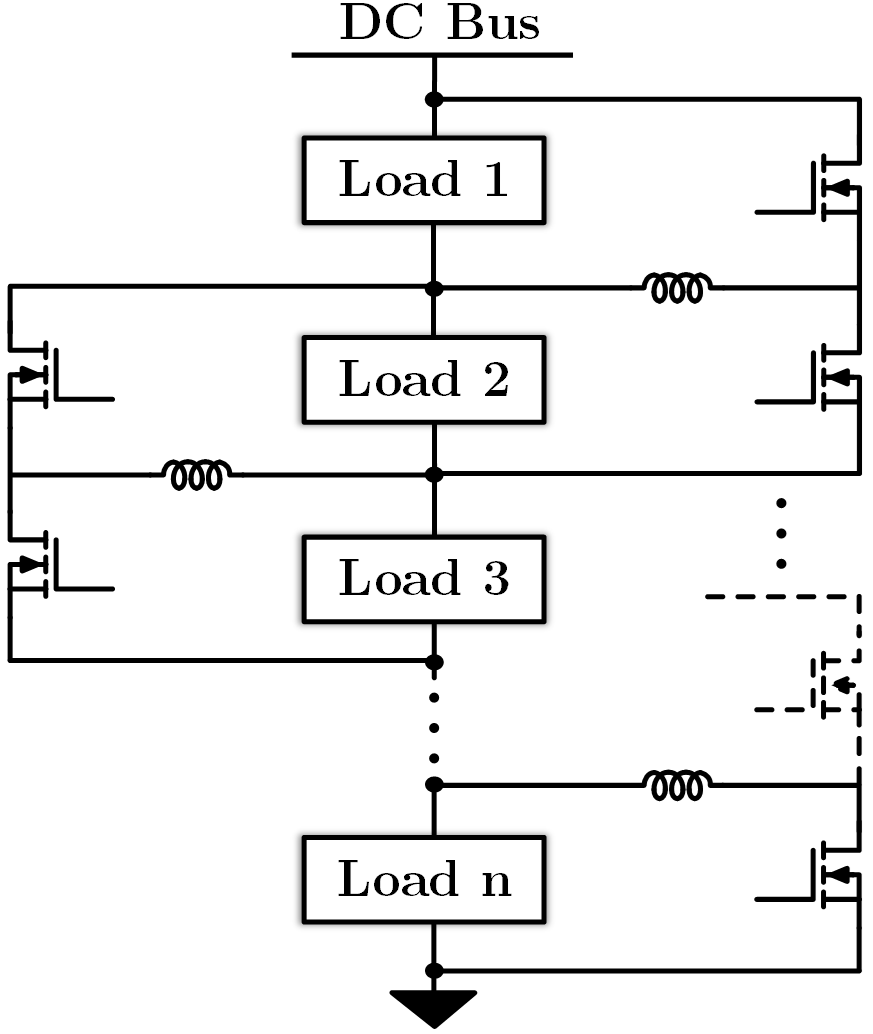}}
		\vspace{-10pt}
		\caption{Example DPP topologies: (a) ac fully-coupled DPP~\cite{Kutkut94}; (b) dc fully-coupled DPP~\cite{Schmidt93}; (c) ladder DPP with dual-active-bridge cells ~\cite{Divan91}; (d) ladder DPP with buck-boost cells ~\cite{Brainard95}. There are many different ways of implementing these topologies.}
		\label{fig:topology}
	\end{center}
\end{figure}

\begin{figure}[t]
\begin{center}
	\vspace{-10pt}
	\subfigure[]{\includegraphics[height=0.175\textheight]{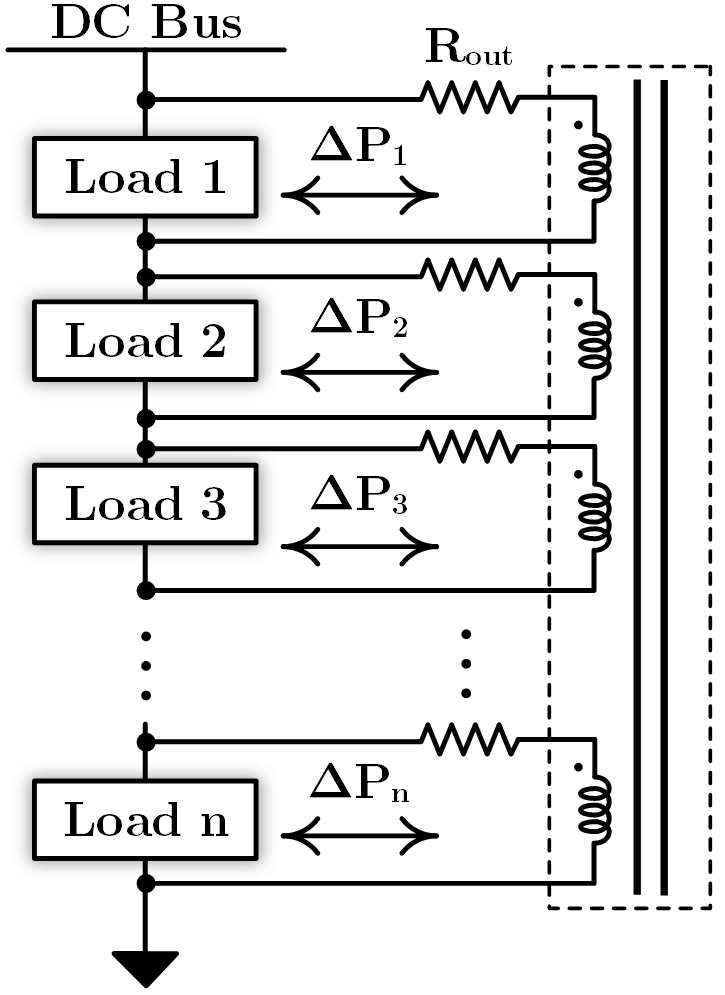}}
	\hspace{0pt}
	\subfigure[]{\includegraphics[height=0.175\textheight]{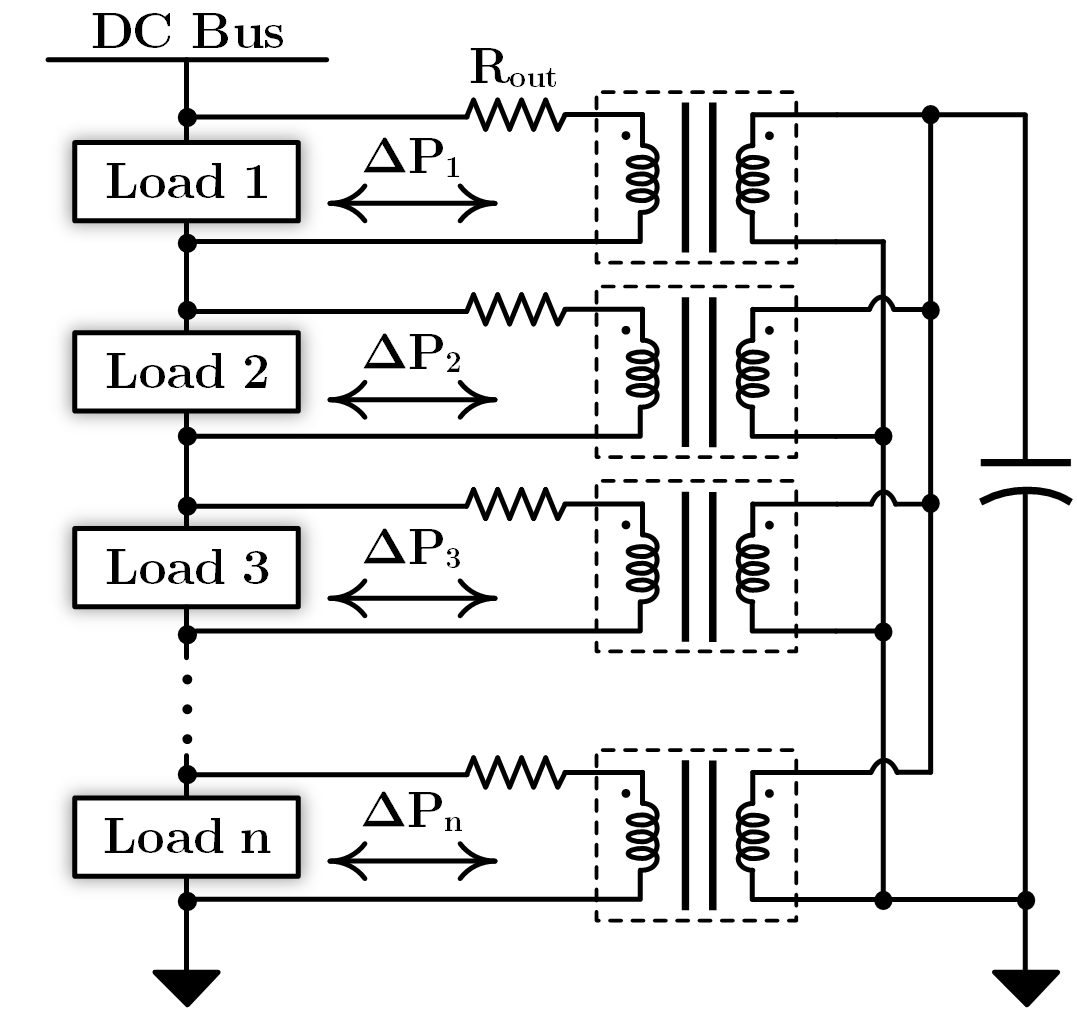}}
	\hspace{0pt}
	\subfigure[]{\includegraphics[height=0.175\textheight]{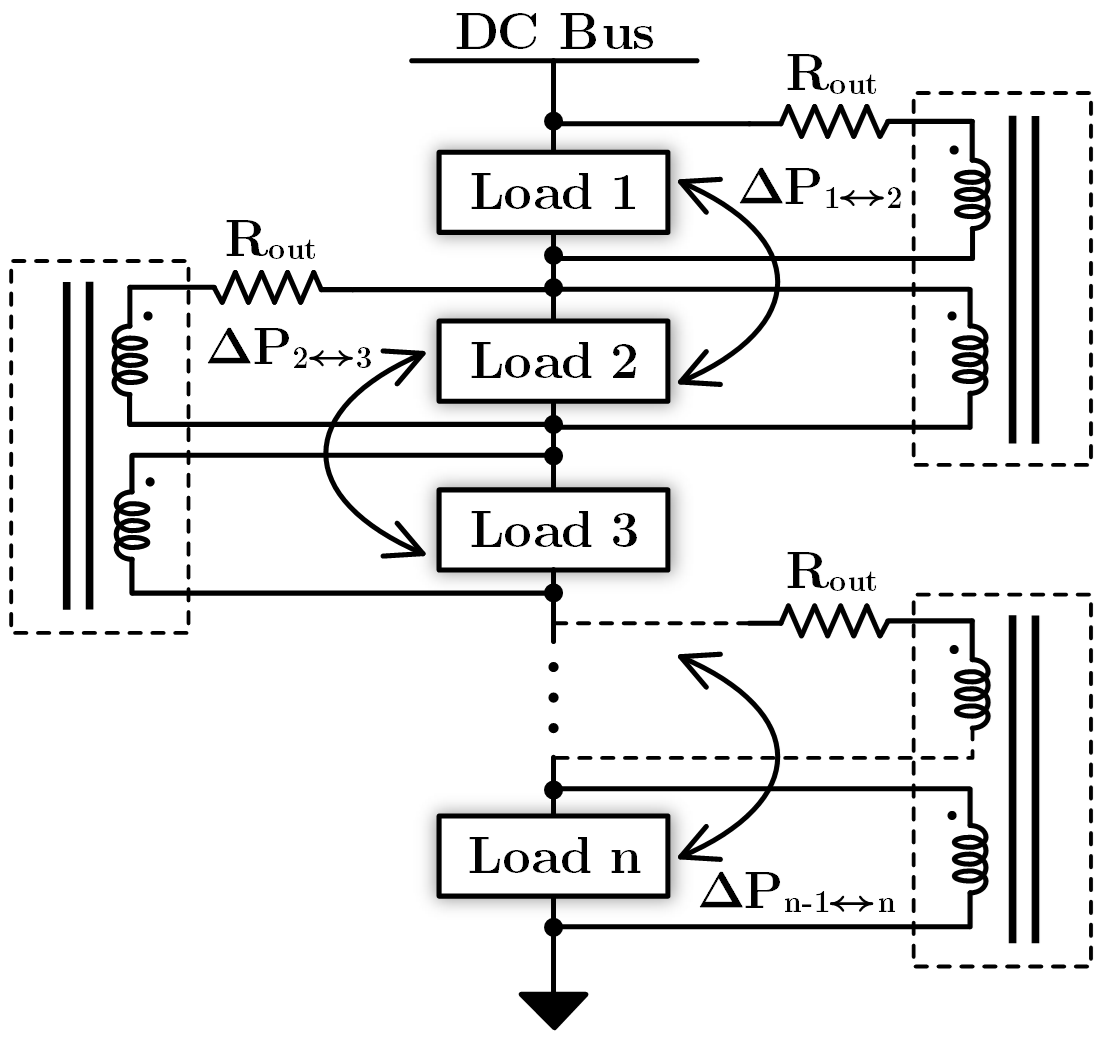}}
	\hspace{2pt}
	\subfigure[]{\includegraphics[height=0.175\textheight]{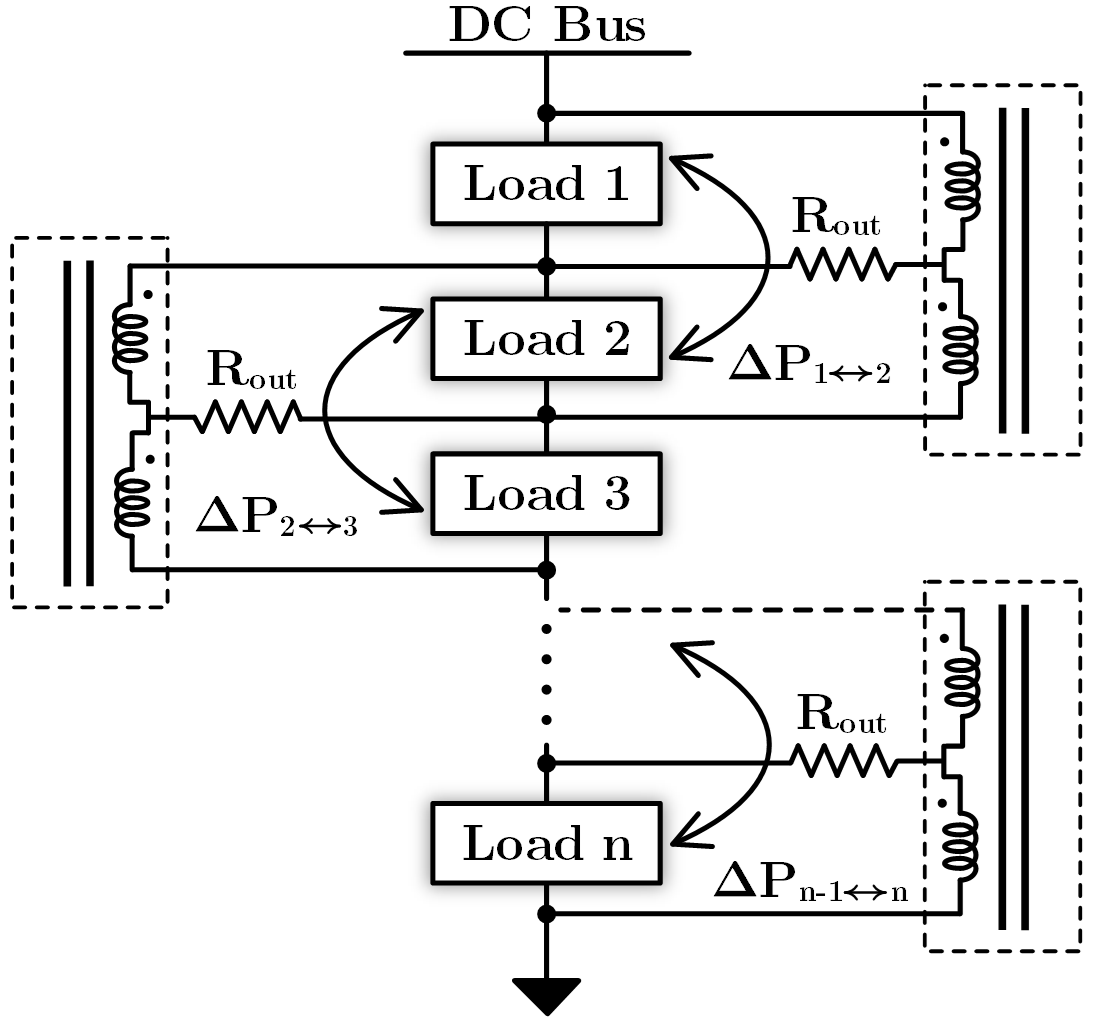}}
	\vspace{-10pt}
	\caption{Simplified models of DPP topologies with the converter modeled as an ideal transformer and an output resistance: (a) ac fully-coupled DPP; (b) dc fully-coupled DPP; (c) ladder DPP with dual-active-bridge cells; (d) ladder DPP with buck-boost cells.}
	\label{fig:lossmodel}
	\vspace{-20pt}
\end{center}
\end{figure}

Fig.~\ref{fig:topology} shows several example DPP topologies classified into two categories: 1) fully-coupled DPP, where there is a direct power flow path between two arbitrary voltage domains; and 2) ladder DPP, where the power is processed by numerous standalone dc-dc converters. Fig.~\ref{fig:topology}a$\sim$\ref{fig:topology}b are fully-coupled DPP topologies, and Fig.~\ref{fig:topology}c$\sim$\ref{fig:topology}d are ladder DPP topologies. Similar analysis will be extended to other DPP topologies in the final paper. DPP solutions are typically applied to systems with stochastic loads. We develop a generalized analysis framework for DPP systems with series-stacked voltage domains and stochastic loads, and compare their performance against an N:1 converter. 

The DPP system in Fig.~\ref{fig:DPP4Array} has $N$ series-stacked voltage domains, each comprising $M$ loads connected in parallel. The voltage of each voltage domain is $V_0$. Assume that the instantaneous power of the $j^{th}$ load in the $i^{th}$ voltage domain is $P_{ij}(t)$. All $P_{ij}(t)$'s are {\it independent and identically distributed} (i.i.d) random variables. The total power consumed by the $i^{th}$ voltage domain is the summation of the $M$ random variables: $P_i(t)= P_{i1}(t)+ P_{i2}(t)+...+ P_{iM}(t)$. A DPP converter manipulates the power flow and balances the series-stacked voltage domains. The instantaneous differential power of the $i^{th}$ voltage domain is the difference between its power and the average power of all domains:
\begin{equation}
\Delta P_{i}(t) = \dfrac{P_1(t) + P_2(t) +...+ P_N(t)}{N} - P_i(t) = \overline{P}(t) - P_i(t).
\label{eq:diffPower}
\end{equation}

\noindent $\Delta P_{i}(t)$ is the minimum power to be processed by the DPP converter. The conduction loss of the DPP converter is directly related to $\Delta P_{i}(t)$. Here we derive the conduction loss of example DPP topologies as a function of $\Delta P_{i}(t)$. 

\begin{easylist}
	\ListProperties(Hide=100, Hang=true, Progressive=0ex, Style*=$\bullet$)
	& ~\textbf{Fully-Coupled DPP Converter}: A fully-coupled DPP topology can be modeled as an $N$-port network with all ports connected to an $N$-winding ideal transformer of uniform turns ratio as shown in Fig.~\ref{fig:lossmodel}a (which is functionally equivalent to Fig.~\ref{fig:lossmodel}b). The conduction loss of the DPP converter is captured by an output resistance $R_{out}$ as labeled in Fig.~\ref{fig:lossmodel}, which is assumed to be identical for all ports. In a fully-coupled DPP converter, the $i^{th}$ port is processing a differential power of $\Delta P_{i}(t)$, and the total conduction loss of the full $N$-port DPP system is:
	\begin{equation}
	P_{loss}(t) = R_{out}\sum_{i=1}^{N}\Delta I_{i}(t)^2 = R_{out}\sum_{i=1}^{N}\left(\frac{\Delta P_i(t)}{V_0} \right)^2 = \frac{R_{out}}{V_0^2} \sum_{i=1}^{N}\left(P_i(t)-\overline{P}(t) \right)^2.
	\label{eq:DirectDPPLoss}
	\end{equation}
	 $P_{loss}(t)$ reflects the variance of an independently repeated sampling experiment of the random variable $P_{i}(t)$. Its statistical expectation, the average power loss of the DPP system over a long enough period, is:
	\begin{equation}
	\mathbb{E}[P_{loss}(t)] = \frac{R_{out}}{V_0^2}\times \mathbb{E}\left[\sum_{i=1}^{n}(P_i(t)-\overline{P}(t))^2 \right] = M(N-1)\frac{R_{out}}{V_0^2}\sigma^2(P_{ij}(t))   \Rightarrow \underbrace{\mathcal{S}(MN\sigma^2)}_{scaling~factor},
	\label{eq:DirectDPPAvgLoss}
	\end{equation}
	where $\sigma^2(P_{ij}(t))$ is the variance of $P_{ij}$. We use the symbol $\mathcal{S}(\bullet)$ to represent the performance scaling factor of a DPP system, which illustrates the growth rate of the loss as the dimension of the DPP system increases. Eq.~(\ref{eq:DirectDPPAvgLoss}) indicates that the performance scaling factor of an $M\times N$ fully-coupled DPP system is $\mathcal{S}(MN\sigma^2)$. \uwave{The expected loss of a fully-coupled DPP converter is only determined by the variance of each individual stochastic load, and scales linearly with $M$ and $N$. The expected loss is independent from the total load power of the system.}
	
	& ~\textbf{Ladder DPP Converter}: In a ladder DPP topology, each DPP unit is a dc-dc converter as shown in Fig.~\ref{fig:topology}c (equivalent as Fig.~\ref{fig:lossmodel}c). Power needs to go through multiple dc-dc converters from one domain to another domain. The differential power that the $i^{th}$ DPP unit needs to process between the $i^{th}$ and the $(i + 1)^{th}$ voltage domains is $\Delta P_{i\leftrightarrow i+1}(t) = \sum_{k=1}^{i} (\overline{P}(t) - P_k(t)) = i\times \overline{P}(t) - \sum_{k = 1}^{i}P_k(t)$, and the total conduction loss is:
	\begin{equation}
	P_{loss}(t) = R_{out}\sum_{i = 1}^{n-1}\Delta I_{i\leftrightarrow i+1}^2 = R_{out}\sum_{i=1}^{n-1}\left(\frac{\Delta P_{i\leftrightarrow i+1}(t)}{V_0} \right)^2 =\frac{R_{out}}{V_0^2} \sum_{i = 1}^{n-1} \left(i\times \overline{P}(t) - \sum_{k = 1}^{i}P_k(t) \right)^2.
	\label{eq:Load2LoadDPPLoss}
	\end{equation} 
     Its expectation, the average power loss of the DPP system over a long enough period, is:
	\begin{equation}
	\mathbb{E}[P_{loss}(t)] = \frac{R_{out}}{V_0^2} \times \mathbb{E}\left[ \sum_{i = 1}^{n-1} \left(i\times \overline{P}(t) - \sum_{k = 1}^{i}P_k(t) \right)^2 \right] = \frac{M(N - 1)(N + 1)}{6}\frac{R_{out}}{V_0^2}\sigma^2(P_{ij}(t))  \Rightarrow \underbrace{\mathcal{S}(MN^2\sigma^2)}_{scaling~factor}.
	\label{eq:Load2LoadDPPAvgLoss}
	\end{equation}
	The conduction loss of an ladder DPP increases linearly as $M$ increases, and increases quadratically as $N$ increases, so the performance scaling factor of a ladder DPP system with $M\times N$ stochastic load array is $\mathcal{S}(MN^2\sigma^2)$. \uwave{The expected loss of a ladder DPP topology is only determined by the variance of each individual load. It scales linearly with $M$, and scales quadratically with $N$, and is independent from the total load power of the system.}
	
	& ~\textbf{Conventional N:1 Dc-Dc Converter}: In a conventional N:1 dc-dc converter, the full power of the $M\times N$ load array needs to be processed by the dc-dc converter. Assuming the output resistance of a conventional N:1 dc-dc converter is $R_{out}$, the conduction loss of this converter when processing power for $M\times N$ i.i.d. loads is:
	\begin{equation}
	\mathbb{E}[P_{loss}(t)] = \frac{R_{out}}{V_0^2} \times \mathbb{E}\left[\left( \sum_{i=1}^{n} P_i(t) \right)^2 \right] = \left(MN\sigma^2(P_{ij}(t)) + M^2N^2\mu^2(P_{ij}(t)) \right)\times \frac{R_{out}}{V_0^2} \Rightarrow \underbrace{\mathcal{S}(M^2N^2\mu^2)}_{scaling~factor},
	\label{eq:CascadedAvgLoss}
	\end{equation}
	where the $\mu(P_{ij}(t))$ is the average power of each unit. \uwave{The expected loss of a N:1 dc-dc converter is determined by the square of the total output power and the variance of each load, and scales quadratically with $M$ and $N$}.
\end{easylist}

Eq.~(\ref{eq:DirectDPPAvgLoss}) and (\ref{eq:Load2LoadDPPAvgLoss}) reveal that the average power loss of DPP architectures is independent from the average power $\mu$ but only determined by the variance $\sigma$, validating the fundamental benefit of DPP solutions: a DPP converter only needs to process the differential power. If the load power is uniform without variation, a DPP system is lossless.

\section{\large Performance Limits, SPICE Verification, and Experimental Results}\label{sec:Comparison}
\begin{table}[t]
	\centering
	\footnotesize
	\caption{Output Resistance, Expected Loss, and Scaling Factor of a few DPP Topologies and an N:1 DAB Converter}
	\begin{tabular}{c|c|c|c|c}
		\toprule[1pt]
		&  Topology  & Output Resistance                 &      Expected Loss                             &    Scaling Factor    \\ \hline
		&&&&	\\ [-1.8ex]
		\multicolumn{1}{c|}{\multirow{3}{*}{Fully-Coupled DPP}} & Ac-Coupled &   $\dfrac{8N}{G_{SW}} + \dfrac{4N}{G_{M}}$     &        \multirow{3}{*}{$M(N-1)\sigma^2(P_{ij}(t))\times \dfrac{R_{out}}{V_0^2}$}         & \multirow{3}{*}{$\mathcal{S}(MN\sigma^2)$} \\ [2ex]
		\multicolumn{1}{c|}{}                                   & Dc-Coupled &    $\dfrac{32N}{G_{SW}} + \dfrac{16N}{G_{M}}$      &                                                                                      &                       \\  [2ex] \hline 
		&&&&	\\ [-1.7ex]	\multicolumn{1}{c|}{\multirow{3}{*}{Ladder DPP}}                                        & DAB-cell  & $\dfrac{32N-32}{G_{SW}} + \dfrac{16N-16}{G_{M}}$ &       $\dfrac{M(N - 1)(N + 1)}{6}\sigma^2(P_{ij}(t))\times \dfrac{R_{out}}{V_0^2}$        &       \multirow{3}{*}{$\mathcal{S}(MN^2\sigma^2)$}    \\  [2.3ex] 
		& Buck-Boost-cell & $\dfrac{8N-8}{G_{SW}} + \dfrac{4N-4}{G_{M}}$ & $\dfrac{2M(N - 1)(N + 1)}{3}\sigma^2(P_{ij}(t))\times \dfrac{R_{out}}{V_0^2}$ & \\ [2.2ex]\hline
		&&&&	\\ [-1.8ex] ~	N:1 Converter                                       &    DAB     &     $\dfrac{32}{G_{SW}} + \dfrac{16}{G_{M}}$     & $\left(MN\sigma^2(P_{ij}(t)) + M^2N^2\mu^2(P_{ij}(t)) \right)\times \dfrac{R_{out}}{V_0^2}$ &       $\mathcal{S}(M^2N^2\mu^2)$        \\ \bottomrule[1pt]
	\end{tabular}
	\label{tbl:Comparison}
	\vspace{-20pt}
\end{table}
\begin{figure}[!b]
	\begin{center}
		\subfigure[]{\includegraphics[height=0.201\textheight]{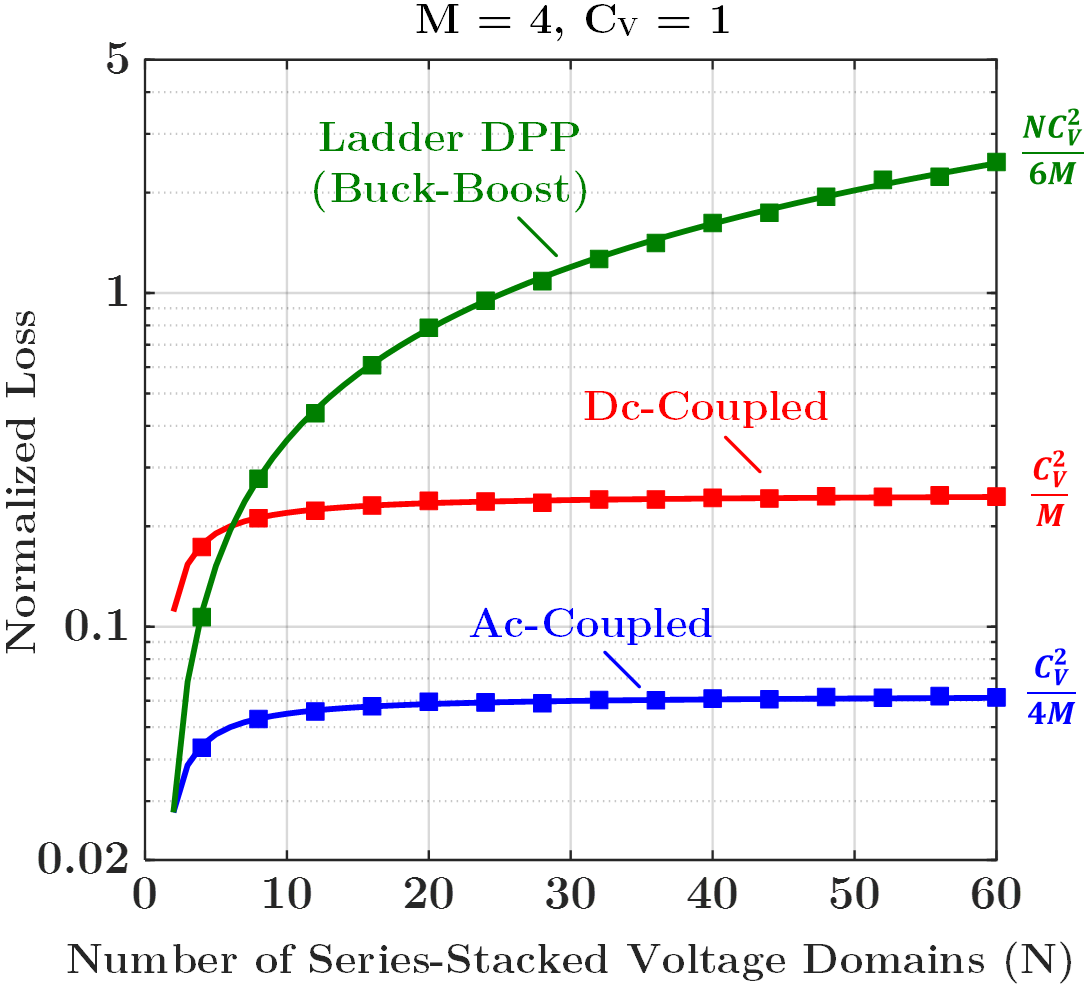}}
		\hspace{0pt}
		\subfigure[]{\includegraphics[height=0.201\textheight]{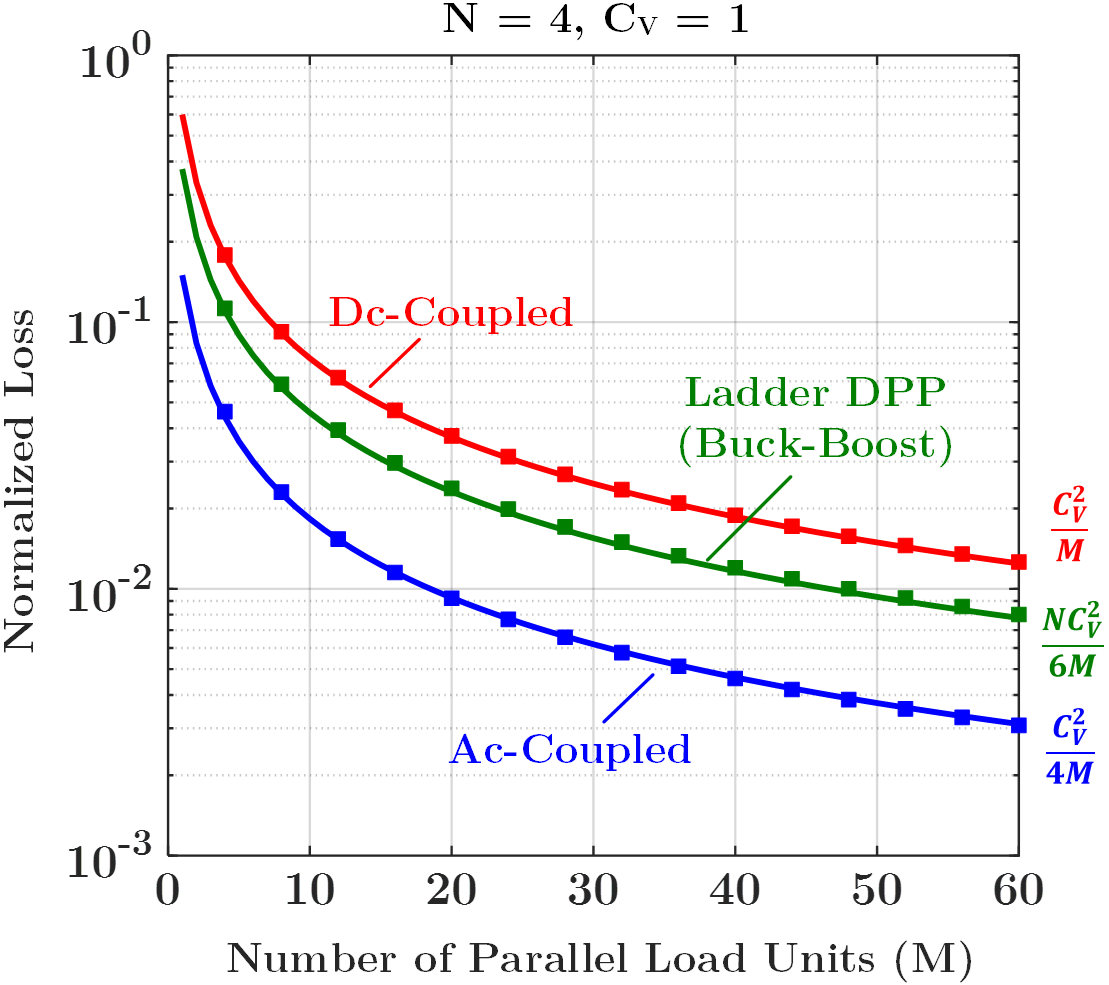}}
		\hspace{0pt}
		\subfigure[]{\includegraphics[height=0.201\textheight]{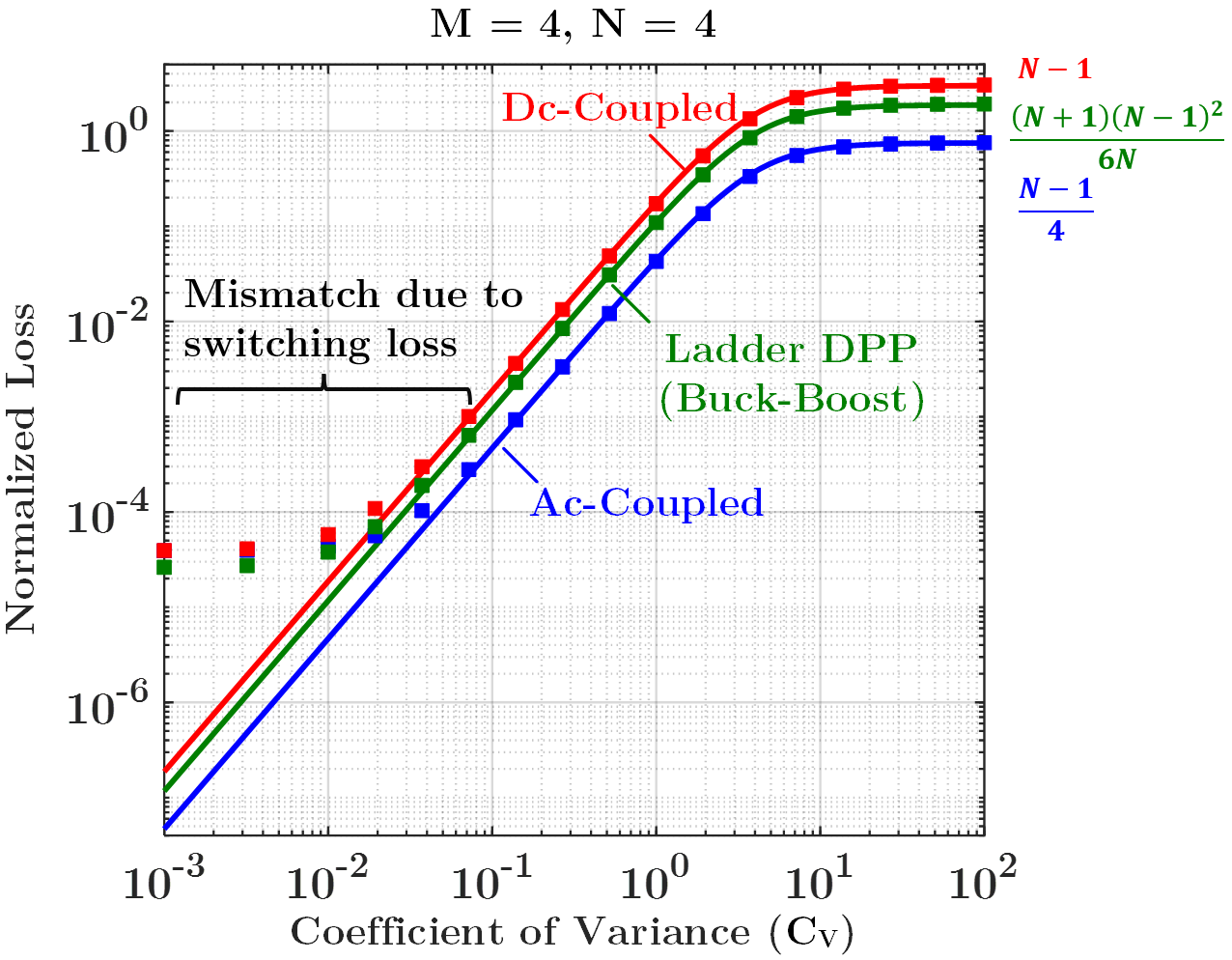}}\hspace{-25pt}
		\vspace{-10pt}
		\caption{Calculated (curve, --) and simulated (dot, $\blacksquare$) normalized loss as functions of: (a) the number of series-stacked voltage domains ($N$); (b) the number of parallel modular loads within one voltage domain ($M$); (c) coefficient of variance ($C_V=\sigma/\mu$) of the loads.}
		\label{fig:ConductionLossRatio}
		\vspace{-20pt}
	\end{center}
\end{figure}

We compare the performance limits of a variety of different differential power processing topologies based on the following assumptions: (a) all topologies have identical semiconductor die area represented by $\sum G_{sw}V_{sw}^2$ ($G_{sw}$ is the switch conductance; $V_{sw}$ is the switch voltage rating); all $\sum G_{sw}V_{sw}^2$'s are normalized to $G_{SW}V_0^2$. (b) all topologies have identical total magnetic window areas represented by $\sum G_{m}$ ($G_{m}$ is the conductance of each winding); all $\sum G_{m}$'s are normalized to $G_{M}$. One way to design an optimal dc-dc converter is to equally allocate the semiconductor die area and transformer window area between input and output ports, and make the design as symmetric as possible. Following the methods in \cite{Makowski95}, we derive the output resistance of the example DPP topologies in Fig.~\ref{fig:topology}, as well as the output resistance of an N:1 dual-active-bridge (DAB) dc-dc converter. The expected losses of DPP topologies and DAB converter are listed in Table~\ref{tbl:Comparison}. Detailed derivations will be provided in the full paper.

We use the ratio between the expected loss of a DPP converter and an N:1 DAB converter (namely Normalized Loss, $\frac{P_{loss,DPP}}{P_{loss,DAB}}$) as a figure-of-merit to evaluate the performance of DPP topologies. The coefficient of variance $C_V$ of $P_{ij}(t)$ is $\sigma/\mu$. The normalized loss of different DPP topologies are shown in Fig.~\ref{fig:ConductionLossRatio}. A lower normalized loss indicates lower loss or smaller volume. Monte Carlo SPICE simulations of DPP circuit models (in Fig.~\ref{fig:lossmodel}) are performed to validate the stochastic loss model. Switching losses are captured by paralleling an equivalent resistance ($1/C_{oss}f_{sw}$) at each port. The simulations are executed 10,000 times to obtain the normalized loss. The simulated normalized loss matches precisely with the calculated results from the theoretical derivation (Table~\ref{tbl:Comparison}).

\begin{wrapfigure}{r}{0.4\textwidth}
	\vspace{-10pt}
	\begin{center}
		\includegraphics[width=0.4\textwidth]{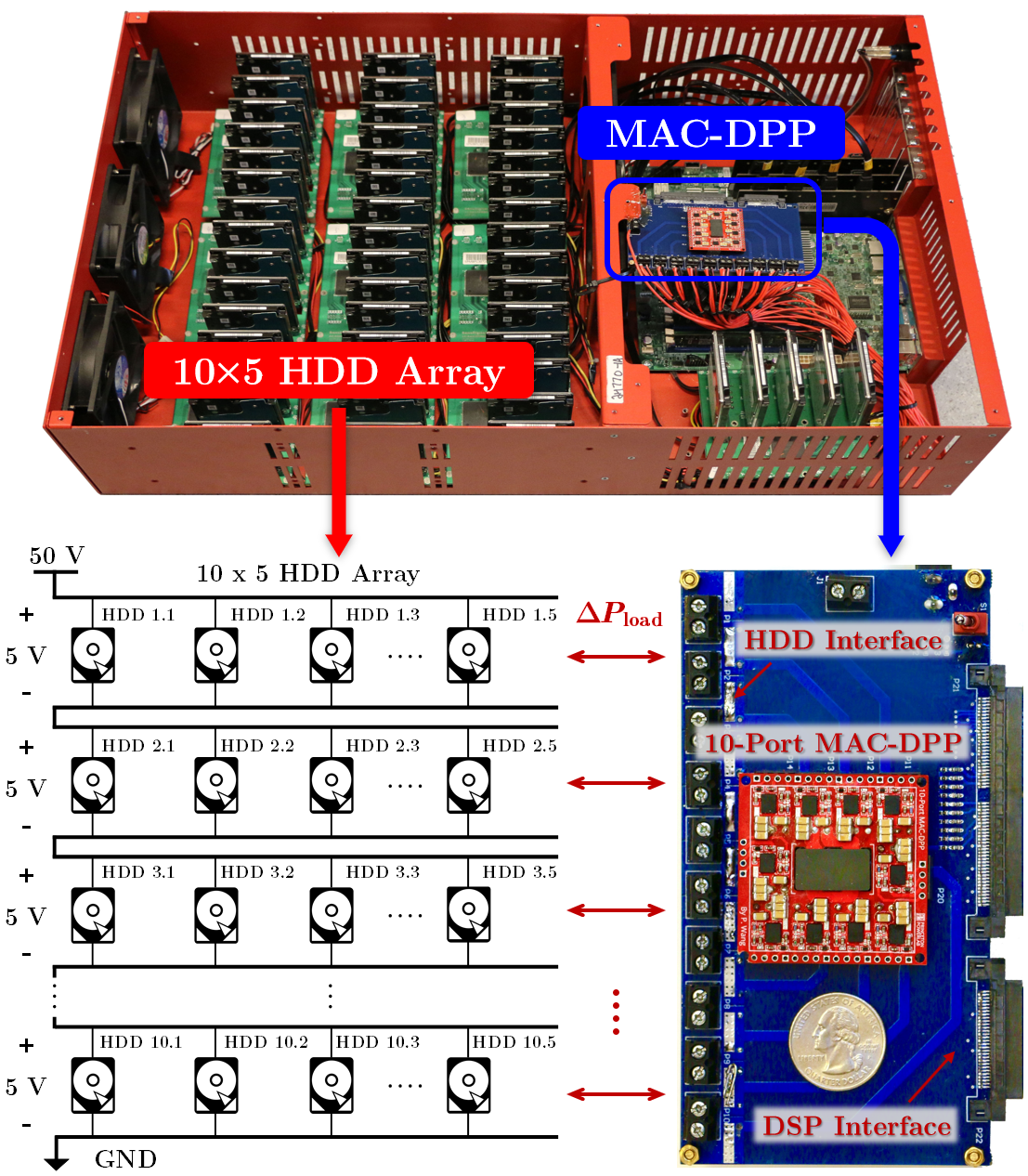}
		\caption{A 50-HDD storage server supported by a 10-port MAC-DPP prototype with 450~W power rating and over 99.5\% peak system efficiency.}
		\label{fig:Testbench}
	\vspace{-25pt}
	\end{center}
\end{wrapfigure}
Fig.~\ref{fig:ConductionLossRatio}a and \ref{fig:ConductionLossRatio}b show the normalized loss of different DPP topologies as $N$ or $M$ increases. As $N$ increases, the normalized loss of the fully-coupled DPP topologies (blue and red) converges to an upper limit, while the normalized loss of the ladder DPP topology (green) keep increasing. This graph provides quantitative design insights for DPP architectures. For example, the normalized loss of the ac fully-coupled DPP topology will converge at $\frac{C_V^2}{4M}$ as $N$ increases. With $M=4$ and $N\ge2$, if $C_V=1$, the normalized loss of an ac fully-coupled DPP converter is always lower than 1/16, indicating over 16x loss reduction from a N:1 dc-dc converter. Fig.~\ref{fig:ConductionLossRatio}c shows the normalized losses of different DPP topologies as $C_V$ changes. When $C_V$ increases, the DPP converter needs to process more differential power, so the normalized losses of all topologies will increase, but they will all converge at an upper limit. This is because the conduction loss of a N:1 converter converter when $C_V$ is large is dominated by $MN\sigma^2$, scaling at the same rate as that of DPP converters. The mismatch at lower $C_V$ range is caused by the switching loss which is not captured by the stochastic model. Fig.~\ref{fig:ConductionLossRatio} reveals that the ac-coupled DPP is the most efficient among all selected DPP topologies. Fig.~\ref{fig:Testbench} shows the picture of a 10-port Multiport-Ac-Coupled DPP (MAC-DPP) converter powering a 10$\times$5 hard-disk-drive (HDD) server. A LabVIEW platform was built to monitor the system efficiency in real time. The 450~W MAC-DPP system reached a peak efficiency of 99.5\% with $>$630~W/in$^3$ power density. Extended experimental results of the storage server with stochastic loads will be presented in the final paper.

\section{\large Conclusions} \label{sec:Conclusions}
This paper reveals the performance limits of differential power processing (DPP). A stochastic model is developed to evaluate the performance limits of DPP topologies as the dimension ($M$, $N$), expectation ($\mu$) and variance ($\sigma$) of modular load array scales up. The performance limits of many DPP topologies are derived and compared, providing useful design guidelines for implementing DPP systems. The analytical framework is verified by SPICE simulations. Extended theoretical derivations and experimental results will be presented in the final paper.

\vspace{-10pt}
\singlespacing
\bibliographystyle{IEEEtran.BTS}

\end{document}